# Large Language Model as An Operator: An Experience-Driven Solution for Day-Ahead Volt/Var Schedule in Distribution Networks

Xu Yang, Chenhui Lin, *Senior Member, IEEE*
*State Key Laboratory of Power Systems*
*Department of Electrical Engineering, Tsinghua University*
Beijing, China
yangxuthu@163.com, linchenhui@tsinghua.edu.cn

Licheng Sha, Liping Yang
*State Grid Beijing Electric Power Company*
Beijing, China
421773717@qq.com, ylpsimple@126.com

Shuzhou Wu, Xichen Tian
*Beijing Key Laboratory of Research and System Evaluation of Power Dispatching Automation Technology (China Electric Power Research Institute)*
Beijing, China
wsz285285@gmail.com, tianxichen2020@163.com

Haotian Liu, *Member, IEEE*, Wenchuan Wu, *Fellow, IEEE*
*State Key Laboratory of Power Systems*
*Department of Electrical Engineering, Tsinghua University*
Beijing, China
liuhaotian@tsinghua.edu.cn, wuwench@tsinghua.edu.cn

***Abstract*—With the advanced reasoning, contextual understanding, and information synthesis capabilities of large language models (LLMs), a novel paradigm emerges for the autonomous generation of dispatch strategies in modern power systems. In this paper, we propose an LLM-based experience-driven day-ahead Volt/Var schedule solution for distribution networks, which enables the self-evolution of LLM agent's strategies through the collaboration and interaction of multiple modules—specifically, experience storage, experience retrieval, experience generation, and experience modification. The experience storage module archives historical operational records and decisions, while the retrieval module selects relevant past cases according to current forecasting conditions. The LLM agent then leverages these retrieved experiences to generate new, context-aware decisions for current situation, which are subsequently refined by the modification module to realize self-evolution of the dispatch policy. Comprehensive experimental results validate the effectiveness of the proposed method and highlight the applicability of LLMs in power system dispatch problems facing incomplete information.**

## I. Introduction

In recent years, large language models (LLMs) have undergone rapid development, marked by substantial increases in model scale, training corpus volume, and architectural sophistication. Modern LLMs exhibit powerful natural language processing and information synthesis capabilities, enabling them to comprehend and reason over heterogeneous knowledge sources. These attributes not only diminish the operational barrier faced by end users but also alleviate the cognitive burden placed on expert engineers [1]. Consequently, there is a growing trend integrating LLMs into power system applications, such as simulation setup [2], power flow calculation [3], document analysis [4], load forecasting [5], and grid model generation [6].

In the context of power system dispatch problems, some pioneering studies have already proposed novel LLM-based dispatch approaches. For instance, LLMs have been leveraged for problem modeling and code generation, translating natural-language dispatch requirements into executable optimization code, which is subsequently solved by commercial solvers [7]. Alternatively, LLMs have also been integrated with reinforcement learning (RL), where the upper-level LLMs analyze grid codes or dispatch requests to provide semantic guidance or structured reward shaping, thereby facilitating the RL training process [8], [9]. In both paradigms, the LLMs act as a high-level reasoning engine that interfaces human intent with computational workflows. In summary, existing LLM-based dispatch approaches fundamentally operate under the assumption of a complete information scenario, wherein the LLMs serve as intelligent coordinators that assist in or invoke model-driven and data-driven tools to derive the final dispatch strategies.

Generally speaking, the aforementioned model-driven and data-driven tools typically rely on access to extensive and highly granular information to formulate reliable dispatch strategies. For example, accurate prediction of voltage profiles necessitates high-resolution forecasts of renewable generation; while precise power flow calculations require detailed, node-

This work was supported by Science and Technology Project of State Grid Corporation of China "Research on Intelligent Situation Awareness and Optimal Decision-Making Technologies for Large-Scale Urban Power Grid Operation" under Grant 52020125000X-025-ZN.

level load data. In the presence of incomplete or coarse-grained information, these methods may suffer from degraded performance or even complete failure.

However, such comprehensive and fine-grained data are often unavailable or impractical to acquire in real-world operational settings. A representative example is the distribution network day-ahead Volt/Var schedule problem discussed in this paper [10], where only coarse-grained aggregated forecasts are available, rendering the above mentioned model-driven and data-driven tools infeasible. Therefore, in such scenarios, LLMs should not just act as coordinators that invoke external tools, but should instead harness their inherent reasoning and knowledge integration capabilities to actively support the generation of day-ahead Volt/Var schedule strategies, thereby mitigating the challenges caused by incomplete information.

Drawing from the authors' practical observations, many dispatch strategies in the real-world power systems are not strictly model-driven or data-driven, but experience-driven. Experience-driven approaches synthesize heterogeneous sources of information, such as dispatch regulations, historical operating conditions, simulation results, and heuristic strategies from domain experts into actionable knowledge. This common phenomenon points to a novel direction for the application of LLMs in dispatch problems, i.e., prior dispatch results can be dynamically incorporated into the LLM agent as "experience," thereby supporting the generation of effective strategies under previously unseen conditions.

Based on this idea, this paper proposes an LLM-based experience-driven solution for the day-ahead Volt/Var schedule problem in distribution networks. The proposed method comprises four core modules: experience storage, experience retrieval, experience generation, and experience modification. These modules operate in a coordinated pipeline: The experience storage module maintains a repository of historical operating conditions and corresponding dispatch strategies; The experience retrieval module identifies and retrieves the most similar past cases based on current forecasting conditions; The experience generation module embeds these retrieved experiences into the LLM agent to generate a context-aware day-ahead Volt/Var schedule strategy for current situation; Finally, the experience modification module iteratively refines the generated strategy through feedback from the external environment, thereby realizing self-evolution of the LLM agent's dispatch policy. Once the experience storage is sufficiently populated, the LLM agent can efficiently generate high-quality Volt/Var schedules in an autonomous manner. Comprehensive experiments demonstrate the effectiveness of the proposed method. To the best of the authors' knowledge, this work presents the first experience-driven framework that leverages LLMs for the direct generation of dispatch strategies in power systems, thereby establishing a new paradigm for LLM applications in the field of power system dispatch.

## II. Methods

This paper focuses on the day-ahead Volt/Var schedule problem in distribution networks with high photovoltaic (PV) penetration. Owing to the intermittent nature of PV generation and variable load patterns, such networks are susceptible to voltage violations during periods of high PV generation or peak load, which motivates the need for proactive day-ahead voltage regulation. Controllable devices available for this purpose include the on-load tap changer (OLTC) at the substation and shunt capacitors (SCs) distributed throughout the network. In the day-ahead stage, operators are given only coarse-grained, hourly forecasts of the total regional load and PV generation for the upcoming day. Based on this limited information, they must determine the OLTC tap positions and SC switching actions over the 24-hour horizon. Objective of the problem is to minimize voltage deviations and violations across the network, while ensuring that the total number of OLTC and SC switching operations does not exceed their respective allowable counts.

To address this problem, our proposed method comprises four modules: experience storage, experience retrieval, experience generation, and experience modification. Their collaborative workflow and interactions are illustrated in Fig. 1.

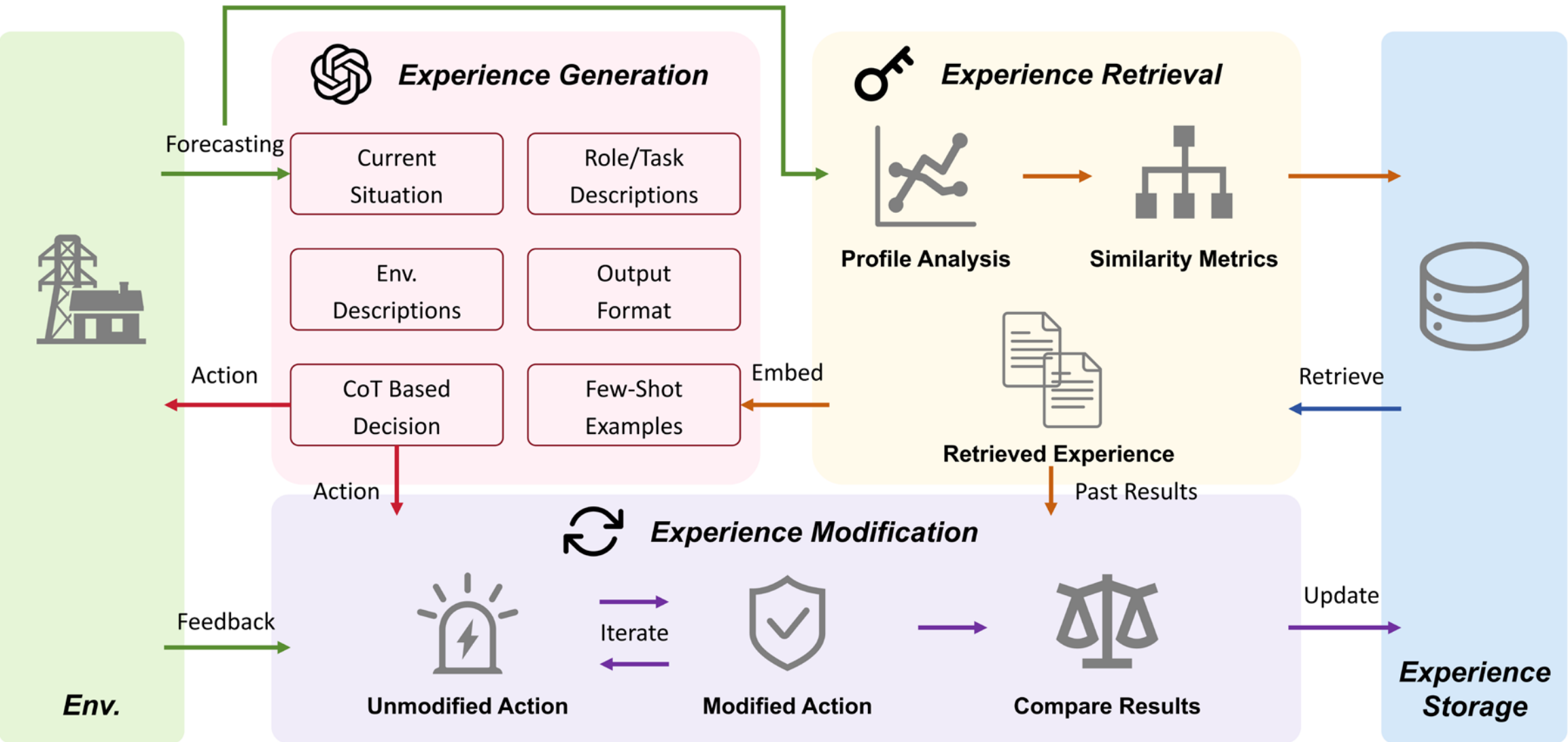

**Fig. 1.** Scheme of the proposed LLM-based experience-driven day-ahead Volt/Var schedule solution.

### A. Experience Storage

The experience storage module is responsible for storing historical operating conditions, dispatch decisions, and their corresponding outcomes, which is initialized by $K$ typical experiences $\{e_i\}_{i=1}^{K}$. In this paper, we propose a structured representation of an experience $e_i$, which comprises the following components:

1) Context of $e_i$: This component outlines the operational context of experience $e_i$, given by the day-ahead hourly forecasts of aggregated load $\left[Load_{i,t}\right]_{t=1}^{T}$ and PV generation $\left[PV_{i,t}\right]_{t=1}^{T}$. It enables similarity comparison with the current forecasting scenario, allowing the experience retrieval module to identify the most relevant past experiences for strategy generation.

2) Reasoning process: This component introduces the reasoning process of the experience, explaining how the final Volt/Var schedule actions are derived from the forecasting data. It records the logical steps, operational considerations, and heuristic judgments used in decision-making, thereby providing a clear reference for future strategy generation. Including this component is essential to enable the LLM agent to understand and replicate expert-like reasoning when facing similar but previously unseen conditions.

3) Final actions: This component records the actual dispatch decisions in experience $e_i$, i.e., the OLTC tap positions $\left[OLTC_{i,t}\right]_{t=1}^{T}$ and SCs switching states $\left\{\left[SC_{i,m,t}\right]_{t=1}^{T}\right\}_{m=1}^{M}$, where $M$ is the number of SCs in the distribution network. These actions serve as concrete examples of effective Volt/Var strategies, enabling the LLM agent to learn and adapt feasible schedules for new scenarios under limited forecasting information.

4) Dispatch results: The final component illustrates the resulting hourly voltage profiles after applying the dispatch actions. It provides feedback on strategy effectiveness, supports performance evaluation, and guides future refinements, which enables continuous improvements from the LLM agent.

It should be noted that the experience storage module can be directly constructed and maintained by human experts. In the absence of expert support, the self-evolution mechanism described below facilitates the completion and proactive updating of the stored experiences.

### B. Experience Retrieval

Just as human experts tend to draw upon the most similar past experiences when encountering new situations, the LLM agent also requires access to the most relevant prior experiences to support current decision-making. Therefore, the role of the experience retrieval module is to identify and retrieve the most similar experiences from the experience storage based on the current scenario.

Considering the characteristics of the Volt/Var schedule problem, where control actions are primarily determined by the timing and magnitude of OLTC tap positions and SC switching states, this paper proposes two similarity metrics: the first is profile similarity, which captures the temporal trends of load and PV generation, and the second is statistical similarity, which characterizes their magnitude information. Calculation of the profile similarity between two experiences $PS(e_i, e_j)$ is the product of PV generation cosine-similarity and the load cosine-similarity:

$$PS(e_i, e_j) = \frac{\sum_{t=1}^{T} PV_{i,t} PV_{j,t}}{\sqrt{\sum_{t=1}^{T} PV_{i,t}^2}\sqrt{\sum_{t=1}^{T} PV_{j,t}^2}} \cdot \frac{\sum_{t=1}^{T} Load_{i,t} Load_{j,t}}{\sqrt{\sum_{t=1}^{T} Load_{i,t}^2}\sqrt{\sum_{t=1}^{T} Load_{j,t}^2}} \quad (1)$$

In the same way, we concatenate the statistical features (including the maximum, minimum, mean value, and standard deviation) into an $N$-dimensional vector, denoted as $\left[Load_{i,n}\right]_{n=1}^{N}$ and $\left[PV_{i,n}\right]_{n=1}^{N}$, based on which the statistical similarity $SS(e_i, e_j)$ is calculated as:

$$SS(e_i, e_j) = \frac{\sum_{n=1}^{N} PV_{i,n} PV_{j,n}}{\sqrt{\sum_{n=1}^{N} PV_{i,n}^2}\sqrt{\sum_{n=1}^{N} PV_{j,n}^2}} \cdot \frac{\sum_{n=1}^{N} Load_{i,n} Load_{j,n}}{\sqrt{\sum_{n=1}^{N} Load_{i,n}^2}\sqrt{\sum_{n=1}^{N} Load_{j,n}^2}} \quad (2)$$

Once the similarity between the current scenario and all stored experiences has been evaluated, the top $k$ experiences with the highest profile and statistical similarity are retrieved from the experience storage and subsequently embedded into the experience generation process.

### C. Experience Generation

As illustrated in Fig. 1, the experience generation module acts as the core decision-making unit. It takes multiple inputs, including retrieved past experiences and hourly load and PV forecasts, and produces Volt/Var schedule actions as output.

To assist the LLM agent in understanding the Volt/Var schedule problem and generating decisions grounded in relevant historical experiences, the prompts designed for this module incorporate the following components:

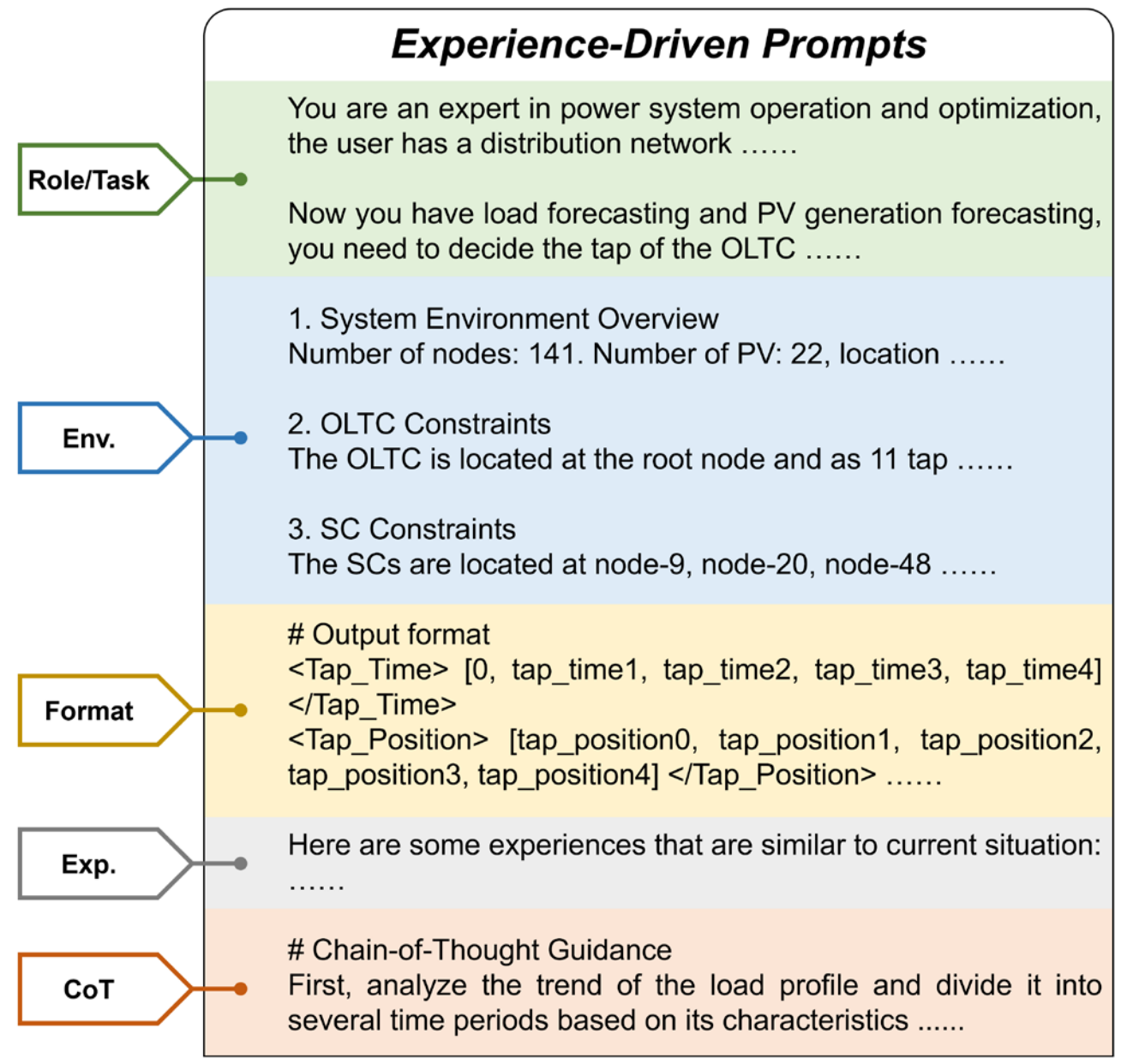

**Fig. 2.** Experience-driven prompts for the experience generation module.

1) Role and task description: This component defines the LLM agent as an expert in power system operation and optimization, tasked with determining day-ahead Volt/Var schedule actions based on the next day's load and PV forecasts.

2) Environment description: This component outlines the distribution network configuration and the Volt/Var schedule problem to be addressed, along with detailed descriptions of the OLTC and SCs and their associated operational constraints.

3) Output format: This component specifies the required output format for the LLM agent. For the day-ahead Volt/Var schedule problem, it requests the action times and magnitudes of the OLTC and SCs to be provided as separate lists.

4) Past experiences: This component integrates the past experiences retrieved by the experience retrieval module to support few-shot learning. Experiences with high profile similarity inform action timing, while those with high statistical similarity guide action magnitude.

5) Chain-of-Thought (CoT) guidance: This component introduces a CoT reasoning framework that begins with analyzing the trend and magnitude of load and PV generation, proceeds to assess potential voltage issues, and concludes with experience-driven decision-making. The CoT structure supports the LLM's reasoning and enables the storage of the corresponding reasoning process.

### *D. Experience Modification*

A major feature of human learning is the ability to improve through feedback from the external environment. Based on this idea, the LLM agent can leverage the interpretability of its actions and its strong reasoning capabilities to iteratively refine the strategies. After the final actions $[OLTC_t]_{t=1}^T$, $\left\{[SC_{m,t}]_{t=1}^T\right\}_{m=1}^M$ are executed in the environment, we first collect the resulting dispatch outcomes, i.e., the hourly voltage statistical data, to form this newly generated experience.

Afterwards, since the actual load and PV generation profiles for the following day become available—along with the observed dispatch results—the strategy can be further refined offline. Specifically, we guide the LLM agent, through multi-round dialogues, to adjust the original actions into refined ones: $[\widetilde{OLTC}_t]_{t=1}^T$, $\left\{[\widetilde{SC}_{m,t}]_{t=1}^T\right\}_{m=1}^M$ based on the collected feedback, and test these refined actions in an offline simulation platform to evaluate whether they yield improved performance compared to the initial actions.

Finally, after the multi-round dialogues, we select the refined strategy with the best performance and compare it against the *k* retrieved experiences. If it outperforms a retrieved experience, it replaces that experience in the storage module, thereby enabling continuous self-evolution of the LLM-based strategy. Considering the length limit, the complete prompts for each module are provided in our online supplementary file [11].

## III. CASE STUDY

In this section, to validate the effectiveness of the proposed method, experiments are conducted on the IEEE 141-bus distribution system [12], which includes 22 PVs with capacity of 1.5MW, 5 SCs each rated at 0.1MVAR, and an OLTC with 11 tap positions from 0.97p.u. to 1.03p.u.. The voltage limitations are set at 0.95p.u. and 1.05p.u..

Test data are randomly sampled from a 3-year historical dataset with 15-minute resolution [13], resulting in 96 power flow cases per day (one episode). However, only hourly aggregated forecasts are available for day-ahead decision-making, rendering conventional model-driven or data-driven optimization approaches infeasible under this information constraint. The network topology and the locations of the controllable devices are illustrated in Fig. 3, and the detailed system configurations are summarized in Table I.

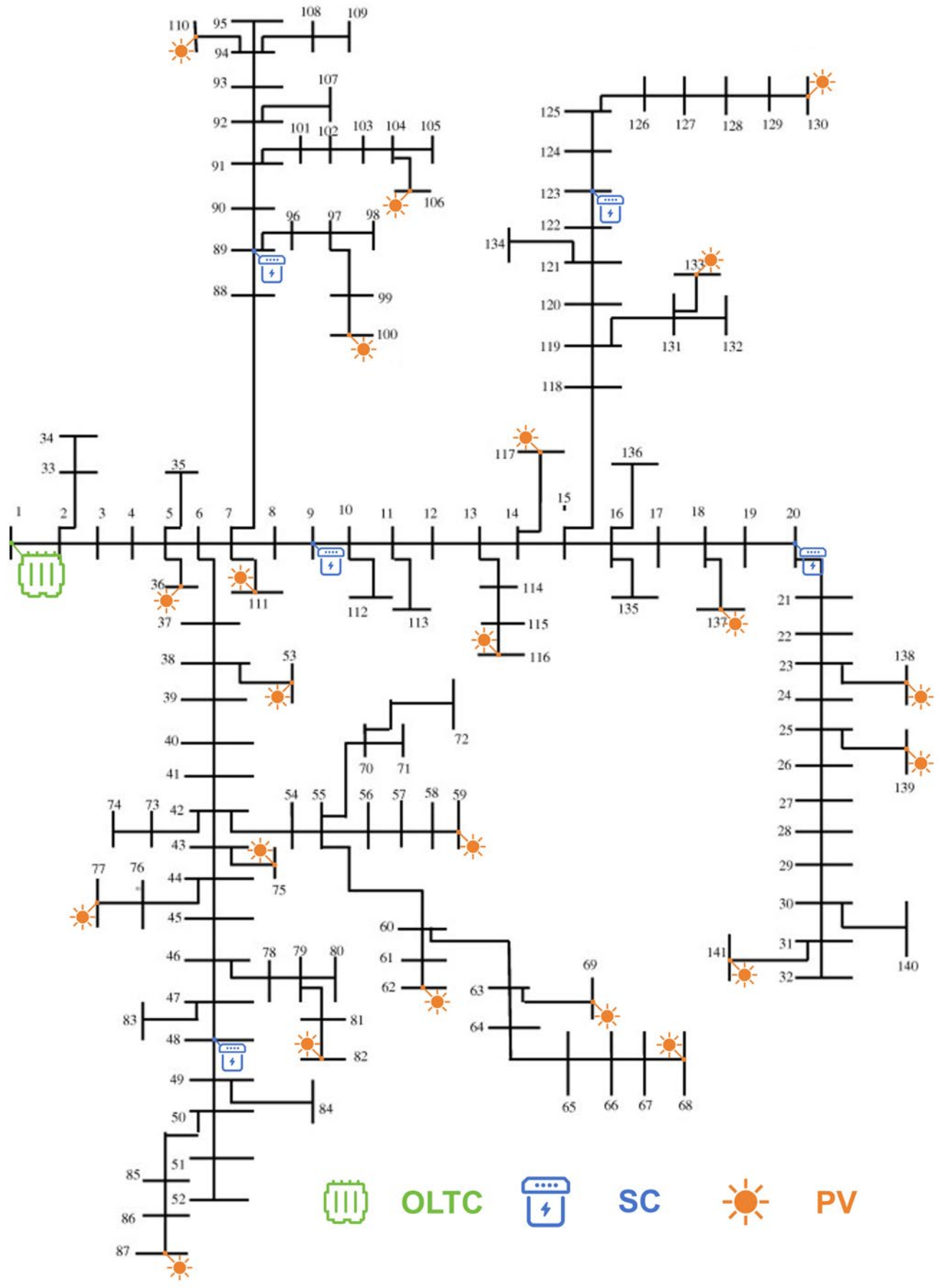

**Fig. 3.** Schematic diagram of the test distribution system.

TABLE I. CONFIGURATIONS OF TEST SYSTEM

| Parameter | Value |
|---|---|
| Number of PVs | 22 |
| Number of SCs | 5 |
| Number of loads | 84 |
| Control interval | 1 hour |
| Data resolution | 15 minutes |
| Voltage limitations | [0.95, 1.05] p.u. |
| PV capacity | 1.5 MW |
| SC capacity | 0.1 MVAR |
| OLTC positions | {-5, -4, …, -1, 0, 1, …, 4, 5} |
| OLTC tap | 0.006 p.u. |
| SC switching limit | 2 |
| OLTC tap change limit | 5 |

Meanwhile, to verify the effectiveness of each component in our proposed framework, we conduct experiments by comparing the proposed method (denoted as Full) against several ablation baselines: NoC (No Control, where OLTC and SCs remain static and uncontrollable), NoE (No Experience, where decisions rely solely on LLM reasoning without retrieved few-shot examples), NoM (No Modification, where the experience modification process is disabled), and NoR (No Reasoning, where the CoT guidance and explicit reasoning steps are omitted). To quantify performance, we also adopt a reward function following the convention in RL, whose value increases as voltage deviations and violations decrease. Table II presents the LLM configurations employed in this paper. More details about the above settings are also provided in [11].

TABLE II. CONFIGURATIONS OF LLMS

| Parameter | Value |
|---|---|
| Model | Qwen-Plus |
| Version | “2025-04-28” |
| Temperature (training) | 0.6 |
| Top-p (training) | 0.7 |
| Temperature (test) | 0.2 |
| Top-p (test) | 0.6 |
| Experience storage size $K$ | 10 |
| Experience retrieval size $k$ | 2 |
| Experience modification rounds | 3 |

Training processes of Full, NoM, and NoR over 3 different random seeds are shown in Fig. 4. As illustrated, the performance of the Full method steadily improves with increasing iterations, demonstrating its capability for self-evolution through interactive and adaptive refinement. In contrast, NoM, which omits the experience modification mechanism, exhibits slower convergence, highlighting the importance of experience modification and updating. Moreover, NoR, which disables the reasoning component, achieves inferior final performance, underscoring the critical role of structured reasoning in guiding effective decision-making.

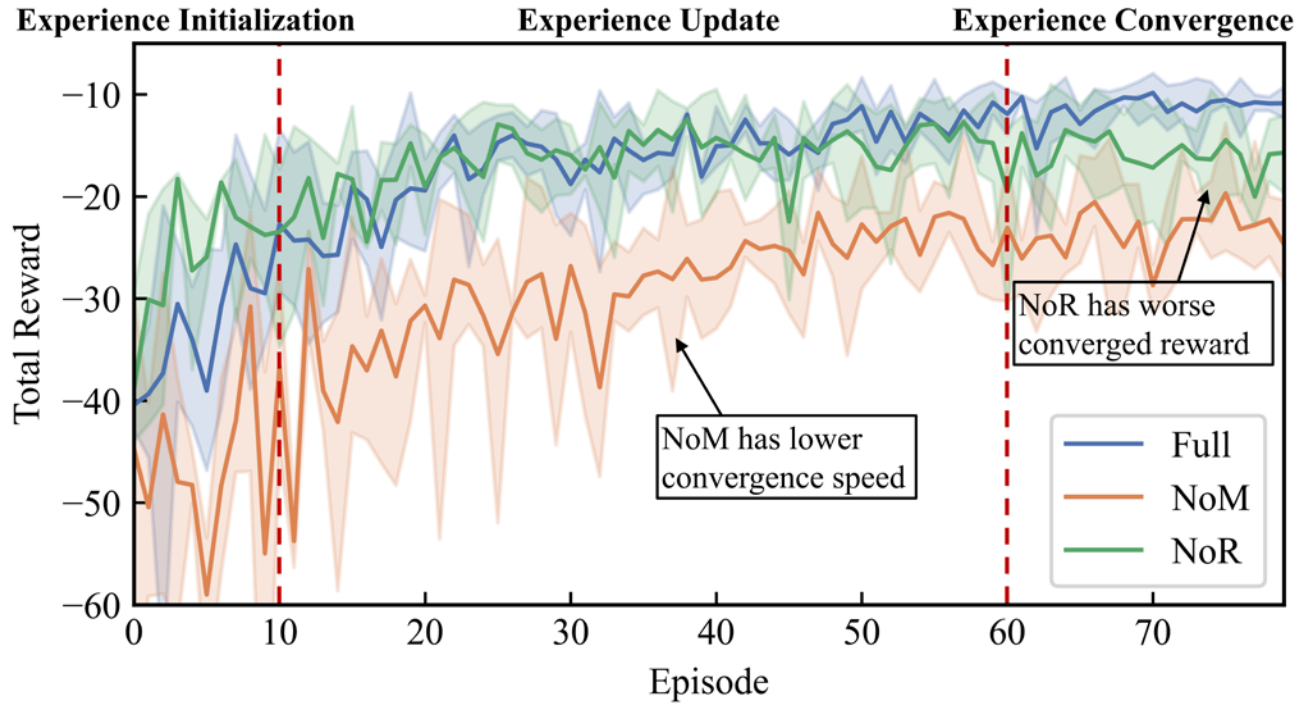

**Fig. 4.** Training process of different methods.

In Table III, we present test results from 30 randomly selected additional episodes, including average reward, voltage deviation, and voltage violation rate for all methods. Compared to NoC, where no control is applied, the proposed Full method achieves substantially improved voltage quality, demonstrating its effectiveness in day-ahead Volt/Var schedule. Moreover, the Full method consistently outperforms all other ablation baselines, confirming the advantages of the LLM-based, experience-driven framework. These results highlight the method’s practical value, particularly in scenarios where limited forecasting information renders conventional model-driven or data-driven approaches ineffective.

TABLE III. TEST PERFORMANCE OF DIFFERENT METHODS

| | **Full** | **NoC** | **NoE** | **NoM** | **NoR** |
|---|---|---|---|---|---|
| **Reward** | -10.7 | -33.4 | -37.2 | -21.3 | -14.9 |
| **Dev.** | 1.08e-02 | 2.26e-02 | 2.10e-02 | 1.66e-02 | 1.27e-02 |
| **Vio. Rate** | 0.139% | 12.5% | 13.3% | 4.97% | 1.84% |

## IV. CONCLUSION

This paper proposes a multi-module collaborative architecture and an associated experience-driven approach, enabling LLMs to develop self-evolving day-ahead Volt/Var schedule strategies with minimal feedback and data. The contributions are twofold: first, it introduces an LLM-based, experience-driven solution for Volt/Var schedule problem; second, it demonstrates the feasibility of directly deploying LLMs in power system dispatch, thereby significantly advancing LLM research and applications in this field.